\documentclass[submission,copyright,creativecommons]{eptcs}
\providecommand{\event}{F-IDE 2018}
\usepackage{underscore}
\usepackage{graphicx}
\usepackage{subcaption}

\usepackage[english]{babel}
\usepackage[T1]{fontenc}
\usepackage{tikz}
\usepackage{hyperref}
\usepackage{xspace}
\usepackage{acronym}
\usepackage{listings}
\usepackage{wasysym}
\usepackage{alloy}

\colorlet{punct}{red!60!black}
\definecolor{background}{HTML}{EEEEEE}
\definecolor{delim}{RGB}{20,105,176}
\colorlet{numb}{magenta!60!black}

\lstdefinelanguage{json}{
    basicstyle=\small\ttfamily,
    showstringspaces=false,
    breaklines=true,
    literate=
     *{0}{{{\color{numb}0}}}{1}
      {1}{{{\color{numb}1}}}{1}
      {2}{{{\color{numb}2}}}{1}
      {3}{{{\color{numb}3}}}{1}
      {4}{{{\color{numb}4}}}{1}
      {5}{{{\color{numb}5}}}{1}
      {6}{{{\color{numb}6}}}{1}
      {7}{{{\color{numb}7}}}{1}
      {8}{{{\color{numb}8}}}{1}
      {9}{{{\color{numb}9}}}{1}
      {:}{{{\color{punct}{:}}}}{1}
      {,}{{{\color{punct}{,}}}}{1}
      {\{}{{{\color{delim}{\{}}}}{1}
      {\}}{{{\color{delim}{\}}}}}{1}
      {[}{{{\color{delim}{[}}}}{1}
      {]}{{{\color{delim}{]}}}}{1},
}

\acrodef{HCI}{Human Computer Interaction}
\acrodef{UI}{User Interface}
\acrodef{VL}{Visual Language}
\newcommand{\etal}{{\it et al.}\xspace}

\title{Improving the Visualization of Alloy Instances}

\author{Rui Couto \and Jos\'{e} C. Campos \and Nuno Macedo \and Alcino Cunha
\institute{Dept. Inform\'atica/University of Minho \& HASLab/INESC TEC, 
Braga, Portugal} }
\def\titlerunning{Improving the Visualisation of Alloy Instances}
\def\authorrunning{R. Couto, J. C. Campos, N. Macedo \& A. Cunha}

\begin{document}

\maketitle

\begin{abstract}
Alloy is a lightweight formal specification language, supported by an IDE, which has proven well-suited for reasoning about software design in early development stages. The IDE provides a visualizer that produces graphical representations of analysis results, which is essential for the proper validation of the model.
Alloy is a rich language but inherently static, so behavior needs to be explicitly encoded and reasoned about.
Even though this is a common scenario, the visualizer presents limitations when dealing with such models.
The main contribution of this paper is a principled approach to generate instance visualizations, which improves the current Alloy Visualizer, focusing on the representation of behavior.
\end{abstract}

\section{Introduction}
\emph{Alloy}~\cite{jackson2012software} is a lightweight formal specification language, supported by an IDE. The IDE consists of a text \emph{Editor} to create the models, an \emph{Analyzer} that allows the user to quickly and automatically analyze desirable properties, and a graphical \emph{Visualizer} to inspect instances (or counter-examples) resulting from the analysis. 
Due to the flexibility of the language, a version of first-order relational logic with an object-oriented flavor, Alloy has proven to be well-suited for modeling and reasoning about software design in early development stages. In particular, it has proven to be specially powerful on design validation, since the Analyzer allows the user to iterate over alternative instances of the model.

The visualizer supports an introspection process, through which developers analyze and understand possible design errors.
Alloy instances are produced in textual format. 
The visualizer takes the instances, and produces textual and graphical representations, the latter being the most used in practice. 
Their appearance can be customized through themes, easing the process of scenario exploration and establishing a common language through which the various stakeholders can communicate~\cite{moreira15orcid}. 
Additionally, it supports focusing (\emph{projecting}) the representation on (\emph{over}) a particular object, thus filtering out all information not related to that object.
This is particularly relevant when analyzing behavioral models. 
Despite Alloy models being inherently static, behavior can be simulated by explicitly modeling state. 
Then, by projecting over states, it is possible to visually inspect behavioral traces, state by state.

The supported visualizations present some limitations, however.
Two issues have been often identified as the most relevant ones in this context. The first is related to the layout algorithms employed~\cite{macedo2016electrum}. The visualizer rigidly organizes objects into rows following an algorithm that is opaque to the user, and although it is possible to move them along the row, it is impossible to move them to a different row. This hinders the capacity to display the elements in a meaningful way. The second is the lack of consistency between different instance representations~\cite{Zave:2015:PCA:2738556.2738583}, as the layout is not preserved. This affects in particular the analysis of different projections of the same instance, further hindering the analysis.

The main contribution of this paper is an approach which improves the introspection process on Alloy traces. We have focused on the representation of transitions between states, specifically in the visualization of changes. 
Drawing inspiration from layout managers~\cite{oracle2017}, as mechanisms that use layout algorithms~\cite{lok2001survey} to visually organize a window, we have developed a flexible approach to specify and manage instance layouts, and the coherent transitioning between projections. The approach is illustrated with an example inspired by the \emph{European Rail Traffic Management System} (ERTMS), creating a more visually appealing representation, and more meaningful transitions between states, than the Alloy Visualizer supports. 

The remainder of this paper is organized as follows. Section 2 introduces the Alloy language, as well as the current challenges identified in the Alloy Visualizer. Section 3 reviews work on improving  the visualizer representations, and Section 4 discusses layout and transition managers.
Section 5 presents the proposed approach, while Section 6 presents its implementation. Section 7 concludes this paper.

\section{Alloy}
\label{sect:alloy}
Alloy is composed of a language and an analyzer supporting software modeling. On the one hand, the language supports the specification of architectures, by describing existing components and their properties. On the other hand, the tool supports the analysis and introspection process, through the analyzer and visualizer, respectively. 

\subsection{The Alloy language}
The structural part of an Alloy model consists of a set of \emph{signatures}, representing sets of uninterpreted atoms, related by \emph{fields} of arbitrary arity. 
Consider the European Railway Traffic Management System (ERTMS) that was the focus of the ABZ 2018 call for case study contributions.
In the very simplified version used herein, we want to model a train track, its virtual partitions (VSS) and trains occupying the track. At any given moment (or state), a train is located in one virtual partition. All the virtual partitions belong to a physical partition, represented by Trackside Train Detection (TTD) sections. Listing~\ref{lst:library} presents the structural components of a possible Alloy model for representing such problem. Signature \texttt{TTD} represents a physical train section. \texttt{VSS} represents a virtual sub section of a track, thus it is associated with exactly one TTD through the binary relation \texttt{ttd}. 

Alloy models are static, so evolution must be explicitly modeled. Here, a signature \texttt{State} models different instants of the evolution of the model, and each field that is expected to evolve must be associated with one such instant. Such is the case of field \texttt{vss}, that relates each \texttt{Train} with exactly one \texttt{VSS} in each state, to express the evolution of the train position. Additional constraints can be imposed through \emph{facts}, expressed in relational logic. These restrict the set of valid instances but do not affect their visualization.

\begin{lstlisting}[language=Alloy,caption=\texttt{ERTMS} Alloy model,label=lst:library]
open util/ordering[State]
open util/ordering[VSS]

sig TTD   { }
sig VSS   { ttd : one TTD }
sig State { }
sig Train { vss : VSS one -> State }
\end{lstlisting}

Formally analyzing Alloy models produces \emph{instances} within a bounded universe specified by the user, which assign to the signatures and fields tuple sets that conform to the model. These can either result from \emph{run} commands and represent \emph{examples} to properties, or from \emph{check} commands and represent \emph{counter-examples} that violate assertions. Both outputs are what allow developers to reason about the specified models and constraints. The Alloy IDE supports the visualization of these instances through the \emph{Alloy Visualizer}. 

\subsection{The Alloy Visualizer}

\begin{figure}
  \centering

  \begin{subfigure}[b]{0.85\textwidth}
    \includegraphics[width=.85\textwidth]{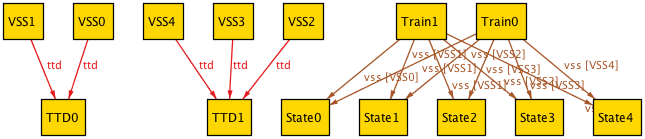}\\
    \caption{An instance of the model}
    \label{fig:libA}
  \end{subfigure}

  \begin{subfigure}[b]{0.45\textwidth}
    \includegraphics[width=.85\textwidth]{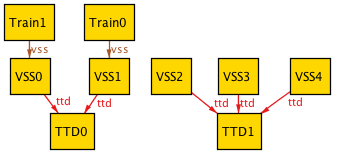}\\
    \caption{Projection over \texttt{State0} (rearranged for clarity)}
    \label{fig:libB}
  \end{subfigure}    
  \begin{subfigure}[b]{0.45\textwidth}
    \includegraphics[width=.85\textwidth]{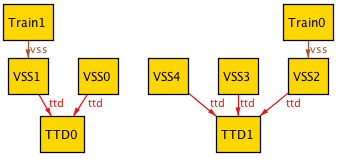}\\
    \caption{Projection over \texttt{State1} (default layout)}
    \label{fig:libC}
  \end{subfigure}
  \caption{Representation of the ERTMS model}
  \label{fig:lib}
\end{figure}

\begin{figure}
  \centering
  \includegraphics[width=.5\textwidth]{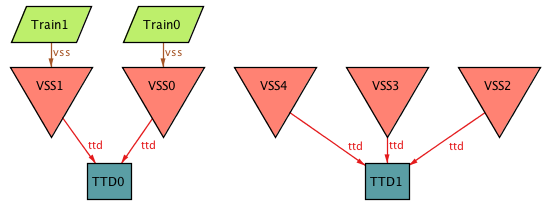}\\
  \caption{Result of the magic layout applied to the instance of Figure~\ref{fig:lib}}
  \label{fig:magic}
\end{figure}

In Alloy graphical representations, the objects are visually depicted by a shape, for instance a rectangle (see Figure~\ref{fig:lib}). These objects are then connected to other objects through arrows, which represent their relations. The visualizer supports customizing the representations through themes. Themes allow developers, for instance, to assign different shapes and colors to signatures and fields, hide or change the way some are represented. Themes can greatly improve the readability of the representations, thus are often used. Another feature provided by the visualizer is the \emph{Magic Layout}~\cite{rayside2007automatic}. This feature automatically creates a theme for a representation, using predefined rules based on the structure of the model, in order to improve it. Figure~\ref{fig:magic} depicts the representation produced by the magic layout feature, when applied to the instance of Figure~\ref{fig:lib}. It is possible to see that the \texttt{Train} objects are represented as green parallelograms, \texttt{VSS} as red triangles and \texttt{TTD} as blue rectangles. Representing objects of the same kind with the same shapes and colors makes it easier to identify them. 

To create a more meaningful representation (without resorting to projections), we could think of moving the train close to its VSS. However, such is not possible with the Alloy Visualizer. 
This greatly impairs the possibility to create more meaningful representations, or simply reorganize elements for a better understanding.

Alloy models with dynamic elements are usually inspected projected over the signature representing the instants of time, \texttt{State} in this model. 
The visualizer will then project out \texttt{State} information and present only information related to a particular \texttt{State} element at a time, which the user can iterate through to interpret the instances.
Figures~\ref{fig:libB} and~\ref{fig:libC} present two projections of the instance at ~\ref{fig:libA}, respectively over \texttt{State0} and \texttt{State1}.

These projections illustrate two problems. First (see Figure~\ref{fig:libC}), \texttt{VSS} instances are shown in reverse order (1 to 0, and 4 to 2).
Second, although it is possible to adjust the layout (this was done in the projection at~\ref{fig:libB}), when iterating over projections those adjustments are lost: as illustrated in the figure, the projection over state \texttt{State1} looses the layout defined for the projection over \texttt{State0}.
The result is that \texttt{Train1} appears at the left most position in both states, although it has moved from \texttt{VSS0} to \texttt{VSS1}, and \texttt{Train0} moves to the right most position from the second leftmost one, although it has only moved one VSS, from \texttt{VSS0} to \texttt{VSS1}. These limitations hinder the understanding of the evolution of trains.

\section{Improving the Alloy Visualizer}
Having identified issues with the default Alloy Visualizer, the next step is to specify how to improve the representations. This corresponds to the identification of a set of requirements, considering our experience with the current visualizer but also related work.

\subsection{Related work}\label{sec:rw}
The first attempt at the improvement of Alloy instances visualization was proposed by Rayside~\etal~\cite{rayside2007automatic}, with the implementation of the \emph{Magic Layout}, a feature now integrated in the Alloy Visualizer, which automatically improves the representation of the instances by automatically generating themes (see Figure~\ref{fig:magic} for an example). Despite the optimizations, issues still persist, for instance with projections, as noted by Zave~\cite{Zave:2015:PCA:2738556.2738583}, as a result of inadequate presentation techniques~\cite{6883044}. Zaman~\etal~\cite{zaman2013improved} propose an approach to improve the Magic Layout features. The authors addressed the highlighting of state changes (on dynamic models), similarities between node layouts of related atoms, and consistency of node positioning between projection states. However, this implementation of the magic layout is not suitable for all kinds of models. For instance, state changes require the objects to have a specific type of behavior. Furthermore, being built on top of the Alloy Visualizer, it has inherited some of its issues, such as limiting the representation of objects to horizontal lines. 

Another type of approach works by optimizing the visualization for specific domains. Gammaitoni and Kelsen presented an approach to represent Alloy instances as state machine representations~\cite{gammaitoni2014domain}. In their work, the authors create representations for Lightning, a sub-language of Alloy. Another example is the work by Bendersky and Galeotti, where the authors create representations of DynAlloy~\cite{bendersky2014dynalloy}, another sub-language of Alloy. The visualizer of Electrum~\cite{macedo2016electrum}, a conservative temporal extension to Alloy, adapts the standard visualizer to support the visualization of (potentially looping) traces.

It can be seen that  efforts have been made  towards the improvement of the Alloy Visualizer. 
However, automatic positioning of elements can only be improved to a certain extent. While the contextual information of the model itself can provide further details, some issues are still to be addressed. It is possible to see that two major groups of approaches exist. On the one hand, there are approaches which deal with extensions to the Alloy language, creating language-specific representations. While these approaches create more expressive representations, they are not suitable for all kinds of Alloy models. 
On the other hand, there are approaches which deal with the base Alloy language. However, issues related to the the visualization of traces, and specifically the changes between projections remain to be addressed by these approaches.

\subsection{Requirements definition}
\label{s:req}

In order to improve the visual representations and support the animation of Alloy models, we have identified a set of requirements which characterize a \emph{good} visualization approach, i.e., that create meaningful representations, and are able to conveniently convey differences between states. These requirements address models with the concept of state, or ordering, and how to support the representation of change over time. The following general requirements were identified, based on past experience and the limitations of related work:

\begin{description}
	\item[Elements rendering (ER)] The first requirement is the capability to render Alloy elements. This corresponds not only to providing a visual representation, but also to the capability to customize such representation. Customizations should support the specification of shapes, colors and images. It should be possible to explore the Alloy signature hierarchy, including the possibility of extending signatures both by extension and inclusion. This requirement is essential in order to create more expressive representations.  

\item[Relations rendering (RR)] In order to support the visual representation of relations between elements, two requirements exist. First, it should be possible to define that an element (e.g. \texttt{Train}) is associated with a property (e.g. color). Second, it should be possible to state that a visual property (e.g. the color of the \texttt{Train}) is the same as the one provided by a relation (e.g.\texttt{vss}). When rendering, the properties of the object will then be determined by its relations. Moreover, relations in Alloy are not limited to the binary case, and may be unary (representing sets of atoms) or have arity higher than 2. These must be properly supported by a visualizer.
	
\item[Projection (Proj)] Not all the information in an Alloy trace has the same relevance. Projections allow the user to view the information related only to a specific element. Projections are useful, for instance, when modeling behavior. In that case, by projecting over state elements, the user can view all the elements and relations in that specific state. However, for the visualization to be useful, the representation of projections over different elements must  be consistent.

\item[Cardinality management (Car)] A variable number of elements can exist in different instances of an Alloy model. Thus, defining properties for a specific element is not a scalable approach.
The ER and RR requirements address defining visual properties for types of objects.
However, they do not address the layout of these objects. In terms of visualization, there is the need for defining relative positions. When the number of elements to render is predefined, it is possible to define absolute positions. However, when the number of elements can vary, properties should specify how a group of elements should be displayed, instead of a single element, i.e., it must be possible to calculate the Layout at runtime. A set of templates should exist in order to display the elements, such as horizontally, vertically or circularly.

\item[Transitions (Tra)] The final requirement is the capability to manage  representations from state to state when using projections. Two sub-requirements can be identified: change and stability management. In short, the changes between consecutive states should be visually highlighted. Animation should be used to deal with change. The term animation is used in a broad sense, meaning from updating the position or color of an element, to adding or removing an element.
Variance management should be done taking into account the cardinality of the elements. It is not viable, for instance, to say that a single object \emph{X} should move from point \emph{A} to \emph{B}, since in another instance the object might not exist, or several objects \emph{X} might exist.  
\end{description}

Table~\ref{tab:comp} presents an analysis of the approaches identified in Section~\ref{sec:rw} against the presented requirements. It is possible to see that while rendering  of elements is common across all the approaches, projection and cardinality management are not found together in the same approach. On top of that, approaches built on top of the Alloy Visualizer inherit its weaknesses regarding Transitions management.
One relevant aspect is that approaches dealing with transitions are domain-specific.
 We propose to improve on existing approaches by addressing Transitions management in the context of a general purpose approach.

\begin{table}[tb]
\centering
\caption{Comparison of presented Alloy visualization techniques}
\label{tab:comp}
\begin{tabular}{lccccc}
\textbf{Author}       & ~~\textbf{ER}~~ & ~~\textbf{RR}~~ & ~\textbf{Proj}~ & ~~\textbf{Car}~~  & ~\textbf{Tra}~ \\

Alloy Visualizer      & $\LEFTcircle$   & $\Circle$       & $\CIRCLE$       &  $\Circle$ & $\Circle$  \\

Rayside~\etal         & $\LEFTcircle$   & $\Circle$       & $\CIRCLE$       &  $\Circle$ & $\Circle$ \\

Zaman~\etal           & $\LEFTcircle$   & $\Circle$       & $\CIRCLE$       &  $\LEFTcircle$ & $\Circle$ \\
\hline
Gammaitoni \& Kelsen  & $\CIRCLE$       & $\Circle$       & $\Circle$       &  $\LEFTcircle$ & $\CIRCLE$ \\

Bendersky \& Galeotti & $\CIRCLE$       & $\CIRCLE$       & $\Circle$       &  $\LEFTcircle$ & $\CIRCLE$
\end{tabular}
\end{table}

\subsection{A new visualizer}
From the above we can conclude that some approaches (the domain-specific ones) support transitions but not projection, while others (the non-domain-specific ones) support  projection but not transitions.
Our proposal is an approach which combines both aspects, allowing for domain-independent representations, while supporting state representations. 

To support transitions in a domain-independent approach, we need to be able to manage how instances and relations are represented and laid out at a global level, and also how to represent changes between states.
Our solution is to assign to each signature/relation managers that will be responsible for laying them out. 
In this way, representations will, on the one hand, be easily configurable and, on the other hand, consistent from  state to state, as intended.

\section{Layout management}
\label{s:lm}
Laying out instances can be compared to laying out user interfaces.
Organizing elements on the screen can be done by resorting to a set of predefined algorithms. 
This supports  creating  dynamic representations, while handling a variable number of elements, instead of rigidly constraining their positioning. Layout managers are mechanisms which support this, assisted by Layout algorithms.

\subsection{Layout managers}
When laying out elements with Layout managers there is a set of elements to display, a space where they will be placed and a set of predefined algorithms to organize the elements in the space. Layout managers provide a clear separation between the information to display, and its representation. On the one hand, there are the elements to render. On the other hand, the configuration of the layouts (i.e., of the algorithms). Proposals for layout managers are available, for instance for Java applications, from Oracle~\cite{oracle2017} and Google~\cite{kellerman2004graphical}. An example is the \emph{Linear} layout (see Figure~\ref{fig:boxlayoutA}), which defines how elements can be displayed sequentially in a row or column. Through layout managers it is possible to define algorithms which control how elements are arranged, as they are added to a layout. 

The graphical organization of \ac{UI} elements is done through hierarchical relations of containment between the elements. 
Each container element will have an associated layout manager.
For instance, an element with \emph{Linear} layout will have child elements, which can themselves be managed by \emph{Linear} layouts.  
Thus, in order to create a user interface resorting to layouts, we need to define a parent layout, and then recursively add child elements and layouts, being the parent a container for the child. This allows, for instance, for constraint based algorithms~\cite{todi2016sketchplore}. Each parent will organize its corresponding space, according to the corresponding layout manager. So, in order to specify \ac{UI}s resorting to layouts, we need to specify: the \emph{parent element}, the \emph{layout manager} (as the algorithm which organizes the child elements), and the \emph{child elements}. Complex layouts are achieved by composing simpler ones. 

In Alloy, an element can be related with another element, being one the child and the other the parent element. 
For instance, in the ERTMS example, a \texttt{Train} is related with a \texttt{VSS}, establishing a hierarchy between those two elements.
However, in laying out Alloy instances, the concept of containment is not suitable. Relations between Alloy instances do not necessarily represent containment.
Instead of a parent container, we need to define a parent anchor, around which elements will be organized. Contrary to layouts, in this case algorithms will be concerned with the space around the anchor elements, instead of the partition of space given to them. In order to create layouts for Alloy instances, we need to define: the \emph{anchor point}, the \emph{elements to organize}, and the \emph{algorithm} to organize them in relation to the anchor.

\subsection{Layout Algorithms}
Layout managers differ between them by the associated algorithm. It is possible to modify the disposition of elements  by changing the algorithm of the associated layout manager. 
Considering that we want to layout Alloy instances around predefined anchor points, algorithms to organize graphs~\cite{herman2000graph} are relevant to define the algorithms required to create Alloy instance layouts. Through graph algorithms, information is automatically organized in a meaningful way. Several approaches exist, as force-directed algorithms and spectral drawing. These algorithms present some more standard features, such as geometrical data representations (e.g. circular, linear, tree), and some more complex representations, for instance oriented by weights~\cite{kobourov2012spring}. 

Another research area dealing with representation of structured information is Information Visualization (Infovis), which focuses on the presentation and exploration of information. 
As above, the processes are assisted by layout algorithms.  Existing approaches resort to abstractions and simplifications, to create better representations~\cite{mazza2009introduction}.  However, in Infovis, a bigger emphasis is put on data representation, rather than understanding the meaning of the data itself~\cite{heer2005prefuse}. 

\subsection{Transition Managers}
A relevant aspect in the representation of Alloy traces, is how differences between consecutive projections are shown. GUI layout managers are usually not concerned with such aspects of the representations, since its not their purpose to manage change. 
Even when considering \emph{plastic} \ac{UI}s~\cite{Thevenin1999PlasticityOU} the focus is on supporting adaptation, not on the transitioning between different \ac{UI} versions.
Differences between projections can be shown in different ways, such as moving elements, or changing their color. 
As above, these properties can also be specified though (transition) managers, which define how a change should be represented.

Layout managers handle how elements should be laid out, managing a variable number of elements. 
However, when an element changes from one layout to another, its behavior (i.e. how the transition is performed) is not specified by the layout manager. Thus, an approach is required to specify how transitions should be processed, i.e., which (transition) algorithm to use. 
Hence, similarly to layouts, transitions between projections can be managed according to a predefined algorithm.  
While layout managers specify how the elements should be organized, transition managers specify how their properties should be adjusted when processing their changes from state to state.

\subsection{Summary}
Despite the relevance of the Alloy Visualizer and its improvements, for instance with the \emph{Magic Layout}, several limitations are well known. Based on the reasoning put forward above, our proposal is to adopt a three-tiered approach.
First, layout managers provide the required background to represent dynamic data. Second, supporting algorithms describe how to render the elements. Third, transition managers allow us to provide systematic transition representations. The next step is to formalize both kinds of layouts, so that a sound implementation might be produced.

\section{Approach}
We have defined as our goal creating a new Alloy Visualizer which improves on the existing one.
The requirements for this new visualizer have been identified. 
We now first describe the  layout managers proposed to organize content and then discuss transition managers.

\subsection{Layout managers}
Each layout manager is defined by its anchor, position, elements and by  variables specific to that layout manager. 
Next we introduce the set of variables common to the descriptions of all layout managers. Remaining variables will be presented when needed.

\begin{itemize}
  \item $A$ - the base Anchor, in relation to which all other elements will be placed
  \item $E$ - an Element to be positioned
  \item $I$ - the Index of an element
  \item $N$ - the Number of elements to render
  \item $S$ - the available Space where to render the elements
  \item $P$ - a Partition of $S$
\end{itemize}

In the layouts presented below, the position of $E$ does not rely on contextual/domain information. 

\begin{figure}[bt]
  \centering
  \begin{subfigure}[b]{0.35\textwidth}
    \includegraphics[width=.85\textwidth]{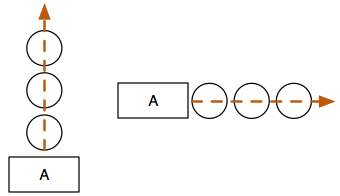}\\
    \caption{North and east configurations of the Linear Layout, with 3 elements each}
    \label{fig:boxlayoutA}
  \end{subfigure}
  \begin{subfigure}[b]{0.20\textwidth}
    \includegraphics[width=.85\textwidth]{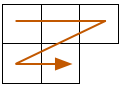}\\
    \caption{Grid Layout Manager}
    \label{fig:boxlayoutB}
  \end{subfigure}
  \begin{subfigure}[b]{0.25\textwidth}
    \includegraphics[width=.85\textwidth]{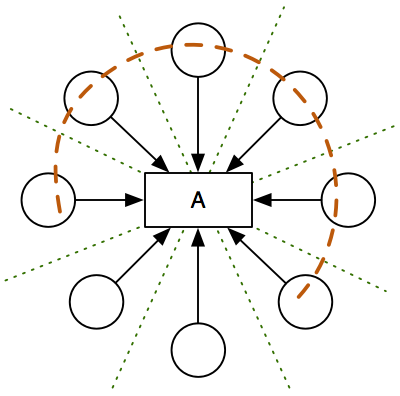}\\
    \caption{Circular Layout Manager}
    \label{fig:boxlayoutC}
  \end{subfigure}
  
  \begin{subfigure}[b]{0.35\textwidth}
    \includegraphics[width=.85\textwidth]{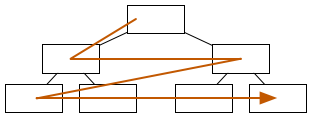}\\
    \caption{Tree Layout Manager}
    \label{fig:boxlayoutD}
  \end{subfigure}
  \begin{subfigure}[b]{0.35\textwidth}
    \includegraphics[width=.85\textwidth]{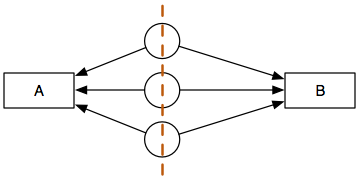}\\
    \caption{Magnet Layout Manager with three elements and two anchors}
    \label{fig:boxlayoutE}
  \end{subfigure}
  \caption{The layout managers}
  \label{fig:boxlayout}
\end{figure}

\paragraph{\bf Linear Layout}
The objective of this layout manager (c.f. Figure~\ref{fig:boxlayoutA}) is to provide a linear rendering of a variable number of elements. Given a direction $D$ (either \emph{north}, \emph{south}, \emph{east} or \emph{west}), the first element is placed next to the anchor, and following elements are aligned with that element (according to $D$). The box layout organizes the visual elements either vertically or horizontally. Elements can also have an order, defining their position. An initial relative position (c.f. \emph{A}) is required to render the first element, as well as its orientation, either horizontal or vertical.

In this layout manager, the available space $S$ needs to be divided into $N$ partitions. Each $E$ is then placed in a partition. Positions can be calculated as follows:
$P = \frac{S}{N}$ where $S = \sum{P}$

The position $Pos$ of a given element is given by:

$Pos(E,I) = if(I==0) \rightarrow (Pos(A) + P) elseif(I>0) \rightarrow (Pos(E,I-1)) + P$

Every time an $E$ is added/removed, $S$ needs to be adjusted to the new $N$, and the element put in the same direction.

\paragraph{\bf Grid Layout}
The Grid layout manager (c.f. Figure~\ref{fig:boxlayoutB}) organizes the elements sequentially, in a bidimensional grid. The first element is placed  to the right of the anchor point. Following elements are placed to the right of the previous one, until filling that row. When a row is complete, the next element is placed in the next row, and so on. Elements can have an order, defining their position in the grid. 

The requirement for this layout is the specification of the \emph{number} of columns $NC$. Elements are added from left to right (until reaching the number of columns), and top to bottom (depending on $N$).
Also, $S$ needs to be divided into $NC$ columns, and the number of rows $NR$ is calculated depending on $N$. The elements are then placed in the appropriate position.

$NC$ is user defined.
$NR = \lceil \frac{N}{NC} \rceil $ where $S = NR \times NC$ 

When an $E$ is added, if the elements count in the bottom row is smaller than $NC$, then no adjustments need to be done. If the count is equal to $NC$, then a new row needs to be added, and $NR$ needs to be adjusted.

\paragraph{\bf Circular Layout}
The circular layout manager (c.f. Figure~\ref{fig:boxlayoutC}) organizes elements circularly, around the anchor point ($A$). The circular space around the center can be divided into as many sections as the elements to be added, and the elements placed in these sections. The order of the elements can be disregarded, since a circular layout does not necessarily have a start and end position. 

The center of a circular layout manager is the $A$ position, around which elements will be placed. $S$ is divided into $N$ partitions, which correspond to a slice $SL$ of a circle. The elements are then placed in the corresponding partitions.

$SL = \frac{360}{N}$ where $S = \sum{SL}$ 

In this layout, the elements are placed along a circular line. Each time an $E$ is added, the number of $SL$ will increase, and their size decrease. Thus, each time $N$ changes, $SL$ will be updated.

\paragraph{\bf Tree Layout}
The tree layout manager (c.f. Figure~\ref{fig:boxlayoutD}) has the objective of organizing elements in a tree structure. Given an anchor element, subsequent elements are placed as its children. Each element has a predefined number of children $NC$, and once an element is full, additional elements will be added as children of the next non full element. Thus, a number of children per element must be defined. 

The tree layout has one parameter, specifying the number of children $NC$ each node should have. The $E$ are then divided in $L$ layers. The process starts by rendering one $E$ in a $L$, and then adding $NC$ $E$s to that node, in another $L$. Next, for each $E$, the process is repeated. When the first $E$ is added, it is placed aligned horizontally with its parent, and directly below it (i.e., next $L$). If no parent exists, it is placed at the center top of $S$. When the next $E$ is added in the same $L$, the $S$ is divided by the $NC$, and the child centered in the $P$.

$\#L=\lceil (\log_{NC}(NC-1)+\log_{NC}(N)\rceil$

When a new element is added, if it exceeds the $NC$ for all nodes in one $L$, a new $L$ is added, and the element added to it. Otherwise, the element is added as child of an available parent $E$.

\paragraph{\bf Magnet Layout}
The Magnet Layout (c.f. Figure~\ref{fig:boxlayoutE}) organizes elements which are attracted to two anchors. Considering an element, in which an attracting force acts towards anchors $A1$ and $A2$, this layout specifies the position for such an element. An element is placed between its two attracting elements, according to its forces. When several elements share the same attraction forces, they are vertically aligned, so that the distances between the attracting elements are proportional. The order of the elements is not considered, since the focus is the distance between the attracting elements. 

Two attracting anchors ($A1$ and $A2$) are required in order to calculate the position of an element. The \emph{forces} ($ST$) related with the elements are also required, in order to calculate the correct position. Elements $E$ should then be placed in between those anchors. The closer their distance to their respective $A$s, the greater that $A$'s $ST$. In order to place the $E$s, the space $S$ between the anchors needs to be divided (according to their strengths) into $D$ partitions. Next, the elements should be positioned taking into account $D$ and the $ST$s.

$E \in \overrightarrow{A1 A2}$. $\#D=\sum{ST}$. Given $P = \frac{\overrightarrow{A1 A2}}{(max(ST1, ST2)-min(ST1,ST2))}$, the position of $E$, $PE$ is given by $PE = \overrightarrow{(A1 P)}$

\paragraph{\bf Random Layout}
The random layout displays the elements in a random position. Each element $E$ is given a random position, inside $S$. Thus, the elements are spread over $S$. The only restriction for this layout is that no two $E$ should overlap.

\paragraph{\bf Absolute Layout}
This layout manager adds all the elements at a specific position on $S$ (e.g. position 0,0). So, no validations or algorithms are applied, and it can be used to let the user manually adjust the positions.

\subsection{Transition managers}
Layout managers create representations, given a set of elements to render and an algorithm. Similarly, transitions can also be systematically created with Transition Managers,  given two sets of elements (i.e., the previous and after states). Two groups of properties can be modified in the  elements, when animating transitions.
\begin{description}
    \item[Rendering properties] correspond to the shape, size, color and position of an object. 
    \item[Visibility properties] correspond to modifying the visibility of an object, by showing/hiding it, according to its existence in a given state. 
\end{description}
Another aspect that can be animated are \textbf{Connections}. Instead of changing the properties of an element, it could remain in the same state, and its connections adjusted instead.

We have drawn inspiration from CSS~\cite{frain2012responsive} in order to define the properties categories. Rendering properties are similar to CSS attributes, which define the visual appearance of an element. Also, CSS transitions and animations allow those properties to be animated. Visibility properties can be compared with the CSS visibility attributes.

Transition managers share similarities with layout managers. Layout managers require a set of elements, and have a space to render them. Also, they require an algorithm in order to arrange the elements. Transition managers require also both a set of elements to update (plus, the information of the current state), and an algorithm to update the new elements. 
Regarding the algorithms, animation managers require the specification of which properties to modify for each signature. 

The most \textbf{Basic} transition manager simply places/removes the elements from the corresponding layout, as they are added/removed. In the case of the projection in Figures~\ref{fig:eg1B} and \ref{fig:eg1C} (see page~\pageref{fig:eg1B}), the object \texttt{Train0} anchor will be shifted from the anchor \texttt{VSS0}, to the anchor \texttt{VSS1}. In this case, the object itself is modified. 

A variation of this manager is the \textbf{Animation} of the elements. Instead of simply removing/adding \texttt{Train0}, the element would be animated, e.g. moving across the screen to the new position. The applicability of transition managers requires certain characteristics in a layout manager in order to be applicable. For instance, if a layout does not rely in positions and proximity, animating the elements position would not make sense.  

\subsection{Summary}
We have presented several layout managers and their specifications. Layout managers satisfy the \emph{Proj} and \emph{Car} requirements presented in Section~\ref{s:req}. Relative positioning of the elements and supporting projections on representations. Layout algorithms deal with the variability on the number of instances. The variability on the number of elements in the representations is also related with the representations of projections in Alloy instances. Additionally, we have discussed the use of transition managers to support transitions between projections, c.f. requirement \emph{Tra}.  

\section{Tool support}
\label{s:ts}
We have developed a tool as an implementation of the proposed approach. The tool supports the representation of Alloy traces (exported as XML files by the Analyzer), and the organization of its elements with layout managers. 
Layout managers supporting absolute, linear, circular and random layouts have been implemented. 
Two different transition managers have also been implemented, one which animates the position of the elements, and another which updates only their connections. 
With this approach we have decoupled the positioning of objects from their transitions. 

The tool requires two inputs in order to create the specifications. First, the trace to render, which contains the instances information. This is automatically produced by the Alloy Analyzer. Second, it requires the layout specification file (in JSON format), to be defined by the users. This file contains all the information required to create the representations, namely association of entities to layout managers and corresponding properties (described in Section~\ref{s:lm}), as well as elements customization. An example of a specification file is presented in Listing~\ref{lst:spec}. 
In the specification is stated that objects of the kind \texttt{VSS} will be layed out according to the \texttt{Linear} layout. Also, the orientation parameter of the \texttt{Linear} layout is set to \texttt{E} (east). Finally, regarding the rendering style, elements will be represented as the \texttt{rail.png} image. 
Combining the trace information with the specification, we are able to create meaningful representations, according to the specified approach. 
\begin{lstlisting}[language=json,caption=Layout specification file,label=lst:spec]
[{"sig":"VSS", "layout":"Linear","base":"ttd","params":["E"],
        "style":{"img":"rail.png","background":"white"}},
{"sig":"Train","layout":"Linear","base":"vss","params":["N"],
       "style":{"img":"train.png","background":"white"}},
{"sig":"TTD", "layout":"Linear","base":"root","params":["S"],
       "style":{"img":"rail.png","background":"white"}}]
\end{lstlisting}

\begin{figure}[tb]
  \centering
  \includegraphics[width=.6\linewidth]{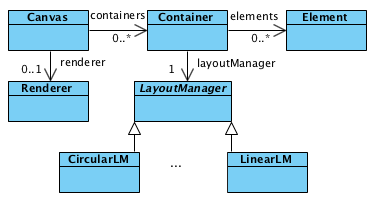}
  \caption{Architecture of the developed tool}
  \label{fig:arch}
\end{figure}

The implementation process  started with the definition the high-level architecture of the tool (see, Figure~\ref{fig:arch}). \texttt{Canvas} is the main class, which represents the canvas to be rendered. It is composed of several \texttt{Container}, which contain the \texttt{Element}s to be rendered. Each \texttt{Container} has also an associated \texttt{LayoutManager}  (e.g. \texttt{CircularLM} or \texttt{LinearLM}), which is responsible for rendering the \texttt{Elements} according to the corresponding algorithms. 
Next, we developed the graphical layer, represented as the \texttt{Renderer}. 
The current \texttt{Renderer} is implemented in Javascript using the Cytoscape\footnote{\url{http://js.cytoscape.org/}, last visited on 2018-04-10.} library in order to create the representations. 
It provides an API to create and manipulate graphical elements, so that the tool can be independent from the concrete graphical library used. 

\begin{figure}[tb]
  \centering
  \begin{subfigure}[b]{0.45\textwidth}
    \includegraphics[width=.85\textwidth]{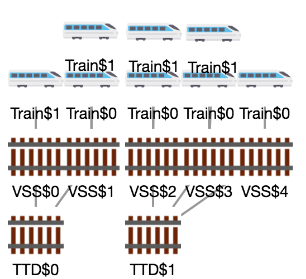}\\
    \caption{Default representation}
    \label{fig:eg1A}
  \end{subfigure}

  \begin{subfigure}[b]{0.45\textwidth}
    \includegraphics[width=.85\textwidth]{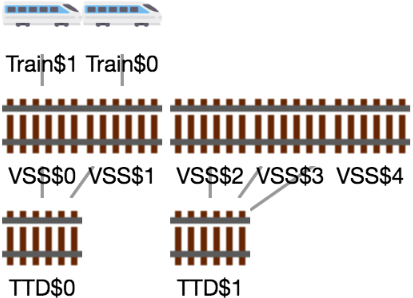}\\
    \caption{Projection over \texttt{State0}}
    \label{fig:eg1B}
  \end{subfigure}
  \begin{subfigure}[b]{0.45\textwidth}
    \includegraphics[width=.85\textwidth]{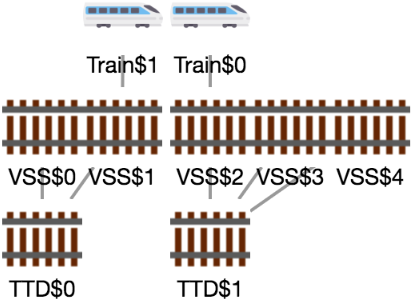}\\
    \caption{Projection over \texttt{State1}}
    \label{fig:eg1C}
  \end{subfigure}
  \caption{An ERTMS model instance and two of its projections}
  \label{fig:eg1}
\end{figure}

Figure~\ref{fig:eg1A} presents the instance from Figure~\ref{fig:libA}, according to the specification presented in Listing~\ref{lst:spec}. As before, there are two instances of \texttt{TTD}, having 2 and 3 associated \texttt{VSS} instances, respectively. Instances of \texttt{VSS} have associated \texttt{Train} instances. It is  possible to see that the visual properties are  applied, with the objects represented as images. 

Figure~\ref{fig:eg1B} and Figure~\ref{fig:eg1C} again depict the projection feature. 
In Figure~\ref{fig:eg1B}, the representation was projected over \texttt{State0}. For each \texttt{VSS} the corresponding train in the state is shown. Advancing to \texttt{State1} (Figure~\ref{fig:eg1C}), the position of the trains is adjusted. Specifically, \texttt{Train0} is moved from \texttt{VSS1} to \texttt{VSS2}, and \texttt{Train1} from \texttt{VSS0} to \texttt{VSS1}. 
When comparing Figure~\ref{fig:eg1} with Figure~\ref{fig:lib}, it is possible to see that \texttt{TTD} elements are now shown in the same horizontal line, below \texttt{VSS} elements. Also, \texttt{VSS} elements are presented sequentially, near the corresponding \texttt{TTD}, and always in the correct order. In the projection, the \texttt{Train} elements are shown above the corresponding \texttt{VSS}. Furthermore, elements are depicted with a corresponding image. 
Regarding the animation, when moving from one state to another state, the trains are moved along the horizontal axis in the screen. They are moved from one \texttt{VSS} to another \texttt{VSS}, making the transitions easier to understand.

\section{Conclusions}
\label{s:c}
This work is integrated into an effort that aims at improving the visualization of Alloy instances. 
The default Alloy Visualizer presents several limitations, namely dealing with variability on the number of instances and with transitions between states. 
To overcome them, we drew inspiration from work on layout managers and supporting algorithms. 
These limitations, and related works supported the formalization of a set of rules which enable the creation of expressive representations. Layout managers are not concerned with the transitions between different configurations. Thus, the concept of transition manager was introduced, as a mechanism which manages the transition of objects in the screen between different layouts. 

We have used a simple model of the \emph{European Rail Traffic Management System} (ERTMS), in order to illustrate the presented approach. With the presented tool, it was possible to create meaningful representations that, by overcoming issues with the existing Alloy Visualizer, demonstrate the viability of the presented approach. 
In particular, the new tool solved issues with the ordering in which VSS segments were layed out and provided consistency across projections. 
Overall,  a more meaningful representation of the instances is provided. 

A set of requirements for a better visualization tool were put forward. 
Namely, Elements Rendering (\emph{ER}), Relations Rendering (\emph{RR}), Projection (\emph{Proj}), Cardinality management (\emph{Car}) and Transitions (\emph{Tra}). 
The paper focused specifically on establishing a framework that supports flexible layout and transitions management.
As a consequence, the tool successfully fulfills requirements \emph{ER}, \emph{Proj} and \emph{Tra}.
Support for requirements \emph{RR} and \emph{Proj} is partial, at the moment. For example, n-ary relations are not currently supported, and representations can only be projected over one signature at a time. 
Additionally, a number of aspects of the Alloy language and its integration with the visualizer were not considered, as they were not key to the set goals.
These aspects need now to be addressed in order to achieve a fully functional visualizer. In the following we set an agenda for future work.

One first, prominent, aspect is support for hierarchies. Alloy specifications support specification of inheritance, where child objects inherit properties of their parents. The inheritance of properties should thus be supported by the tool. 
Still on the language, support for n-ary relations needs to be improved.
Indeed, in Figure~\ref{fig:eg1A} we can see that \texttt{VSS1} is associated with both \texttt{Train1} and \texttt{Train2} but not that this happens in different states.
While the projections clearly show this, better support for representing more than  binary relations needs to be investigated.

Another aspect relates to the integration of the models with the visualization. 
At the moment, the specification of the layouts must be manually created by the developers. 
A better approach will be to annotate the Alloy code, specifying properties for signatures. Those properties can then be used to automatically generate the layout specifications. 

A fourth aspect relates to projections, which are currently not fully compliant with how the Alloy Visualizer handles them. 
To ease the transition from the original visualizer, this aspect also needs to be addressed.

A fifth aspect is related to validation.  
While it can be argued that the flexibility and monotonicity of the representations will make the visualizations easier to understand than the standard ones, end user validation will be helpful in providing further concrete evidence that is the case. 

One advantage of the proposed approach is extensibility, so additional layout and animation managers can be added. 
Finally, while the visualizer is currently a standalone tool, its integration into an IDE is envisaged. 

\subsubsection*{Acknowledgments}
This work is financed by the ERDF - European Regional Development Fund through the Operational Programme for Competitiveness and Internationalisation - COMPETE 2020 Programme and by National Funds through the Portuguese funding agency, FCT - Funda\c{c}\~ao para a Ci\^encia e a Tecnologia, within project POCI-01-0145-FEDER-016826.

\nocite{*}
\bibliographystyle{eptcs}
\bibliography{sigproc}
\end{document}